\documentclass[twocolumn,prl,amsmath,amssymb,showpacs,
superscriptaddress,floatfix]{revtex4}
\usepackage{epsfig,graphicx}
\DeclareMathOperator{\Tr}{Tr}
\DeclareMathOperator{\Sp}{sp} \DeclareMathOperator{\tr}{tr}
\DeclareMathOperator{\diag}{diag}
\newcommand{\Z}{\mathcal Z}

\begin{document}
\title{Energy relaxation in the spin-polarized disordered electron %Fermi
liquid.
}

\author{N.\,M.\,Chtchelkatchev}
\author{I.\,S.\,Burmistrov}
%\email{nms@landau.ac.ru}
\affiliation{L.D.\ Landau Institute for Theoretical Physics,
Russian Academy of Sciences, 117940 Moscow, Russia}
\affiliation{Department of Theoretical Physics, Moscow Institute
of Physics and Technology, 141700 Moscow, Russia}

\begin{abstract}
The energy relaxation in the spin-polarized disordered electron
systems is studied in the diffusive regime. We derived the quantum
kinetic equation in which the kernel of electron-electron
collision integral explicitly depends on the electron
magnetization. As the consequence, the inelastic scattering rate
is found to have non-monotonic dependence on the spin polarization
of the electron system.
\end{abstract}


\maketitle

Solid state mesoscopic electronic systems provide an outstanding
microlaboratory for various experiments on low-temperature
physics. In particular, it allows to study the fundamental
relaxation mechanisms in solids that is important for finding a
way to create advanced cryogenic devices. Last decade large
progress have been achieved in fabrication of devices in which
electron distributions can be controlled and manipulated. For
example, they are electronic refrigerators, thermometers,
radiation detectors, and distribution controlled
transistors~\cite{Giazotto_review}. The main building block of the
devices is a diffusive normal metal or a heavily doped
semiconductor wire connected to massive electrodes acting as
reservoirs. One of the basic goals of the manipulation with
electron distributions is to cool electrons in the diffusive metal
or the wire much below the lattice temperature. In general,
electrons interact among themselves; also they are coupled to
phonons and  to the electromagnetic
environment. Qualitative understanding of the performance of the
devices is based on the ability to solve the quantum kinetic
equation for the electron (spin) density matrix $\hat f$ in which
interactions are taken into account via scattering
integrals~\cite{Schmid,AronovAltshulerTau}. At low temperatures
($T$) which are typical for experiments electron-electron
interaction provides usually the strongest mechanism for energy
relaxation. It can be characterized by the out-scattering rate
($1/\tau_\textrm{out}$) that appears in the kinetic equation
formalism and has the meaning  of the inelastic
electron-electron collisions frequency~\cite{Aronov-Altshuler}. With its
help one can estimate time and length scales at which a
non-equilibrium electron distribution function, e.g., after the
cooling cycle, can be approximated by the Fermi-Dirac distribution
with some effective electron temperature~\cite{Giazotto_review}.

Usually, a magnetic field is used as one of convenient tools for
manipulation with the electron distribution
function~\cite{Anthore2,Giazotto_PRL}. In a weak magnetic field
(the Zeeman energy is much smaller than the Thouless energy) the
collision integral $I[\hat f]$ acquires magnetic field dependence
only due to the shift of electron energy in the distribution
function due to the Zeeman splitting~\cite{Mineev}. However, in
typical experiments~\cite{Huard,Giazotto_review} the Thouless
energy is comparable with the temperature: $T\sim 0.1$ K, and,
therefore, magnetic field above $0.1$T should be treated as
strong.
%The main objective of the present Letter is to
%consider the energy relaxation in diffusive normal metal due to
%electron-electron interaction in the presence of strong magnetic
%field . \textbf{!orbital effects are ignored}
In this regime the magnetic field not only shifts electron energy
in the distribution function but changes the frequency dependence
of the scattering probability such that a kernel of the collision
integral becomes magnetic field dependent. This effect should be
especially important for the disordered electron systems due to
diffusive propagation of electron-hole excitations.

\begin{figure}[t]
  \includegraphics[width=80mm]{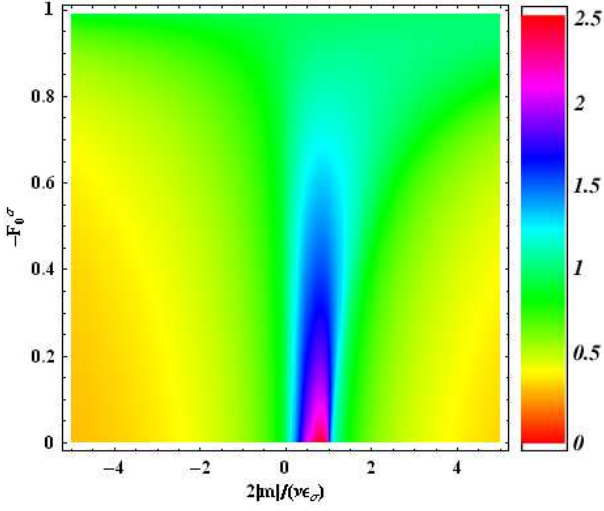}
  \caption{\label{fig1}   (Color online) The color density plot of
  $\tau_\textrm{out}^{(t)}(0)/\tau_\textrm{out}^{(t)}(|\mathbf{m}|)$
  as a function of $F_0^\sigma$ and $2 |\mathbf{m}|/(\nu
  \epsilon_\sigma)$.
%The illustration of the magnetic field
%induced energy-relaxation rate enhance in the dirty metal.
%For example, in Ni$_3$Ga, $F_t=-0.96$; in Pd, $F_t=-0.9$ and in Pt,
%$F_t=-0.5$. The temperature here is supposed to be smaller than the
%Zeeman energy $g\mu_B H$ and $\epsilon_\sigma$ - the energy of a spin-$\sigma$
%electron measured with the respect to the corresponding chemical potential $\mu_\sigma$.
%It follows that the enhance may reach 300$\%$.
}
\end{figure}

In the Letter we consider the energy relaxation due to
electron-electron interaction in disordered electron systems in
the presence of a large Zeeman splitting $ 2 |\mathbf{m}|/\nu$,
where $\nu$ denotes the thermodynamic density of states per one
spin projection. The average spin density $\mathbf{m}$ of the
electron system caused either by the applied magnetic field or by
an appropriate boundary conditions. We derive the quantum kinetic
equation in which the kernel of the inelastic collision integral
explicitly depends on $\mathbf{m}$ (cf.
Eqs.~\eqref{QKE}-\eqref{ColInt}). We find that in the
out-scattering rate which can be written as the sum of
contributions from the singlet ($1/\tau_\textrm{out}^{(s)}$) and
triplet ($1/\tau_\textrm{out}^{(t)}$) channels:
$1/\tau_\textrm{out} = 1/\tau_\textrm{out}^{(s)} +
1/\tau_\textrm{out}^{(t)}(0)+2/\tau_\textrm{out}^{(t)}(|\mathbf{m}|)$,
contributions from the triplet channel with non-zero spin
projection on the $\mathbf{m}$ direction are strongly affected by
the Zeeman splitting. For example, at $T=0$ and in the case of
$d=3$ dimensions we obtain {(we use units $\hbar=c=k_B=1$)}
\begin{equation}
\frac{1}{\tau_\textrm{out}^{(t)}(|\mathbf{m}|)} =
\frac{2^{1/2}}{6\pi^2} \frac{|\epsilon_\sigma|^{3/2}}{\nu D^{3/2}}
\frac{\gamma [(1+\gamma)^{3/2}-1] }{2+\gamma}
\mathrm{F}_\gamma\left (\frac{2 |\mathbf{m}|
}{\nu\epsilon_\sigma}\right ). \label{tauOutd3}
\end{equation}
Here, $\gamma = -F_0^\sigma/(1+F_0^\sigma)$ where
$F_0^\sigma$($F_0^\rho$) stands for the standard Fermi liquid
interaction parameter in the triplet (singlet) channel, $D$
denotes the diffusion coefficient, and $\epsilon_\sigma=\epsilon -
\sigma|\mathbf{m}|/\nu$ where $\epsilon$ is an electron energy
with respect to the Fermi energy $E_F$. The function
$\mathrm{F}_\gamma(z)$ has the following asymptotics
\begin{gather}
\mathrm{F}_\gamma(z) =
  \begin{cases}
    1+ \frac{6+9 \gamma +3(\gamma-2)\sqrt{1+\gamma}}{2(2+\gamma)[(1+\gamma)^{3/2}-1]}z &
    \quad |z|\ll 1 , \\
    \frac{9 \gamma(2+\gamma)}{16[(1+\gamma)^{3/2}-1]}|z|^{-1/2} & \quad |z|\gg 1+\gamma.
  \end{cases}
\end{gather}
Therefore, at zero temperature the
$1/\tau_\textrm{out}^{(t)}(|\mathbf{m}|)$ is the non-monotonic
function of $|\mathbf{m}|$ at fixed quasiparticle energy
$\epsilon_\sigma$. It has the maximum when $2 |\mathbf{m}|/\nu
\sim \epsilon_\sigma$, and is strongly suppressed at larger
$|\mathbf{m}|$ (see Fig.~\ref{fig1}). This non-monotonic
dependence of $1/\tau_\textrm{out}^{(t)}(|\mathbf{m}|)$ on
$|\mathbf{m}|$ is similar to the non-monotonic behavior with
magnetic field for the scattering rate of electron on a magnetic
impurity~\cite{Anthore1,Goppert}.

%%%%%%%%%%%%%%%%%%%%%%%%%%%%%%%%%%%%%%%%%%%%%%%%%%%%%
%%%%%%%%%%%%%%%%%%%%%%%%%%%%%%%%%%%%%%%%%%%%%%%%%%%%%%

The most general method to describe a non-equilibrium low-energy
dynamics in disordered interacting electron systems is the Keldysh
non-linear $\sigma$-model~\cite{Kamenev-Andreev}. In the presence
of non-zero $F_0^\sigma$ as well as $\mathbf{m}$, one can follow
the same derivation of the quantum kinetic equation as given in
Ref.~\cite{Kamenev-Andreev,Aleiner1,Aleiner2} except for
complications that arise from non-commutativity of different
components of plasmon and gauge fields. Below we sketch the
derivation of the quantum kinetic equation briefly and highlight
the points where it differs from one of
Refs.~\cite{Kamenev-Andreev,Aleiner1,Aleiner2}.

We write the grand partition function of the interacting electrons
in a random potential in the coherent state basis: $\Z=\int
D\overline{\psi}D\psi \exp\{iS[\overline{\psi},\psi]\}$, where
$S[\overline{\psi},\psi]=\int_{\cal
C}dt\left\{\int_x\Sp(\overline{\psi}i\partial_t{\psi})-H[\overline{\psi},\psi]\right\}$.
Here, ${\cal C}$ is the Keldysh contour, the symbol $\Sp$ denotes
trace over spin degrees of freedom and $H=H_0+H_{\rm int}$. The
one-particle Hamiltonian $H_0$ involves the parameters with
standard Fermi liquid renormalizations~\cite{AGD}. The interacting
part $H_{\rm int}= \int d\mathbf{r}d\mathbf{r^\prime}\left\{\frac
1
2\hat\rho_{\mathbf{r}}\Gamma_s(\mathbf{r}-\mathbf{r^\prime})\hat\rho_{\mathbf{r}^\prime}+
2 \mathbf{\hat
m}_{\mathbf{r}}\Gamma_t(\mathbf{r}-\mathbf{r^\prime})\mathbf{\hat
m}_{\mathbf{r^\prime}}\right\}$ where
$\Gamma_t(\mathbf{q})=F^{\sigma}_0/(2\nu)$ and
$\Gamma_s(\mathbf{q})=V_0(\mathbf{q})+F^{\rho}_0/(2\nu)$ contains
the long-range part of the Coulomb interaction
$V_0$~\cite{AGD,FinkelsteinReview}. The charge and spin density
operators are given as
$
\hat\rho_\mathbf{r}=\sum_\sigma\bar\psi_\sigma(\mathbf{r}t)\psi_\sigma(\mathbf{r}t)$
and $\mathbf{\hat m}_\mathbf{r}=
\frac{1}{2}\sum_\sigma\bar\psi_\sigma(\mathbf{r}t)\hat{\mathbf{s}}_{\sigma\sigma'}\psi_{\sigma'}(\mathbf{r}t)$,
respectively, where the Pauli-matrices $\hat s^\alpha$,
$\alpha=\overrightarrow{{0,3}}$ act in the spin space.

\begin{figure}[t]
\includegraphics[height=7.5mm]{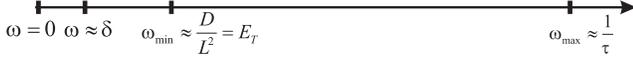}
\caption{``Large'' and ``small'' energy scales of the problem.
Here $\delta$ is the mean level spacing and $L$ is characteristic
system length. The slow in time fields, e.g. $\langle
\Theta^\alpha \rangle$, have the Fourier transforms which are
localized within the energy interval $(\delta,\omega_{\rm min})$.
The Fourier transforms of the fast in time fields, e.g. $\delta
\Theta^\alpha$ that describe the density and magnetization
fluctuations, are inside $(\omega_{\rm min},\omega_{\rm max})$.
Similar diagram can be drawn for the definition of slow and fast
spatial scales. Integrating out fast degrees of freedom in the
partition function, $\Z$, we obtain the kinetic equation for the
Wigner transform of the one-particle density matrix which is slow
in space and time.} \label{fig:scales}
\end{figure}

To derive the non-linear $\sigma$-model we perform standard
steps~\cite{Kamenev-Andreev}: i) we average $\Z$ over the
Gaussian, $\delta$-correlated disorder and then, introduce the
$\tilde Q$-matrix; ii) next, we decouple four-fermion interaction
terms in $H_\textrm{int}$ by the Hubard-Stratonovich
transformation using the vector field $\Theta^\alpha$,
$\alpha=\overrightarrow{{0,3}}$
%[zero component takes part in the decoupling of the singlet interaction
%term and the other components --- the triplet interaction term]
; iii) then, we perform the Keldysh rotation~\cite{Rammer-Smith}
and iv) finally, we integrate out fermion degrees of freedom.
Hence, we obtain
\begin{multline}\label{eq:action_log}
i S[\tilde Q,\Theta]= \Tr\ln\biggl[i\partial_t-\xi+\frac i
{2\tau}\tilde Q+(\Theta^\alpha_j-\phi^\alpha_j) \hat s^\alpha\hat
\gamma_j \biggr]
\\
+\frac i2\int dt d\mathbf{r} [\Theta^\tau
\hat\Gamma^{-1}\hat\sigma_z \Theta]-\frac{\pi\nu}{4\tau}
\Tr(\tilde Q^2),
\end{multline}
where symbol $\Tr$ denotes the trace over the spin and Keldysh
spaces combined with time and space integrations, $\tau$ is the
elastic scattering time. The Pauli matrices $\hat\sigma_z$,
$\hat\gamma_1=\hat\sigma_0$ and $\hat\gamma_2=\hat\sigma_x$
operate in the Keldysh space, $\xi=p^2/2m_e-E_F$ where
$m_e$ is the electron mass,
%$\nu$ is free electron DoS at the Fermi shell;
 and $\hat\Gamma$ is the diagonal matrix in the $4\times4$
$\Theta$-space: $\hat\Gamma=\hat
F/(2\nu)=\diag\{\Gamma_s,\Gamma_t,\Gamma_t,\Gamma_t\}$. We assume
that there are an electric potential $\varphi$ and a static
magnetic field $H$. Then, the classical components
$\phi^{\alpha=0}_1=e\varphi$ and
$\phi^{\alpha>0}_1=-g\mu_B\,H_\alpha$ where $g$ and $\mu_B$ denote
$g$-factor and the Bohr magneton, respectively; the quantum
components $\phi^\alpha_2\equiv 0$. The low-energy description is
valid under the following conditions: $\phi_1^\alpha\tau\ll 1$ and
$T\tau\ll 1\ll E_F\tau$, see also Fig.\ref{fig:scales}. In
addition, we assume $T/E_F\lesssim\tau eH/m_e\ll 1$ and,
therefore, ignore the Cooper channel and orbital effects.

The $\Theta$-field has a nonzero (zero) average for the classical
(quantum) components, [$\langle\ldots\rangle=\int
(\ldots)\exp(iS)$]:
\begin{equation}\label{eq:<Phi>-<Theta>}
\langle\Theta^{\alpha=0}_1\rangle=-\frac{F^{\rho}_0}{2\nu} \rho,
\,\,
\langle\Theta^{\alpha>0}_1\rangle=-\frac{F_0^\sigma}{\nu}\mathbf{m}_\alpha,\,\,
\langle\Theta^{\alpha}_2\rangle = 0
\end{equation}
where $\rho$ is the average charge density. From here onwards, we
omit $V_0(\mathbf{q})$ for a sake of simplicity: it can be always
restored by taking the so-called ``unitary'' limit,
$F_0^\rho\to\infty$~\cite{Aronov-Altshuler}. Next, to improve the
convergence of expansion of the logarithm in
Eq.~\eqref{eq:action_log}, we perform the gauge rotation of the
$Q$-matrix~\cite{Kamenev-Andreev,PruiskenBaranovSkoric}:
\begin{gather}
\tilde Q_{t,t'}(\mathbf{r})  = U(t,\mathbf{r}) Q_{t,t'}(\mathbf{r}) U^{-1}(t',\mathbf{r}),\,\,{Q}=\begin{pmatrix}
  R & K \\
  Z & A \\
\end{pmatrix},\notag \\
U(t,\mathbf{r})=\exp\{i \langle\check k\rangle\}\exp\{i \delta
\check k\}, \,\check k\equiv k^\alpha_i\gamma^{i} \hat s_\alpha.
\end{gather}
%where the ``check'' has the following meaning: $\check k\equiv k^\alpha_i\gamma^{i} \hat s_\alpha$.

In general, the gauge $k$ and plasmon $\Theta$ fields can be
separated into slow, $\langle k\rangle$ and $\langle \Theta
\rangle$, and fast, $\delta k$ and $\delta \Theta$, contributions
as compared to the Thouless energy (see Fig.\ref{fig:scales}).
%E.g, the $\langle k_1^0\rangle$-field performs the
%gauge transformation of electromagnetic potentials.
Physically, the fast components describe
charge and spin fluctuations in the electron system.
%In the same manner we write $\Theta=\langle \Theta\rangle+\delta\Theta$.
%[We'll specify below how $\delta k$ is related to $\delta \Theta$.]
Under assumptions that $E_F \gg 1/\tau\gg T$ and $r_s=e^2/
v_F\lesssim 1$ {where $v_F$ is the Fermi velocity},
%[we are far from the Stoner instability \cite{Aleiner1}] and
we expand the action~\eqref{eq:action_log} around the standard saddle-point $Q=\Lambda$ with $Z=0$,
$R=-A=1$ and arbitrary $K$. In the equilibrium, the Wigner transform
$K_{\epsilon}(\mathbf{r},t)=\int dt^\prime K_{t+t^\prime/2,t-t^\prime/2}(\mathbf{r})(\exp(i\epsilon
t^\prime)$ is equal to
$K^{eq}_{\epsilon}(\mathbf{r},t)= 2\tanh[(\epsilon-\varphi-\partial_t \langle k^\alpha_1\rangle \hat
s^\alpha)/2T]$. As usual, we restrict ourselves to the second order in
$\delta k$, $\delta\Theta$ and
gradients of $Q$ within the low-energy manifold $Q^2=1$.
%----------------------------------------
%----------------------------------------
%----------------------------------------
%----------------------------------------

Then, we find
\begin{multline}\label{eq:S}
iS=i\int dt\int dr[\Theta^\tau \hat \Gamma^{-1}\hat
\sigma_x\Theta+\nu\, b^\tau \hat \sigma_x b]\\-\frac{\pi\nu}
4\left[D\Tr(\partial_r Q)^2+4i\Tr(i\partial_t+\check b) Q\right].
\end{multline}
%where $\partial_r Q= \nabla Q +i[\mathbf{\check g},\,Q]_-$.
%Here $D$ is the diffusion coefficient,
Here, $\partial_r Q =\nabla Q +i[\mathbf{\check g},\,Q]_-$,
$\check b =
U^{-1}(-\check\phi+\check\Theta)U+U^{-1}[i\partial_t,U]_-$  and
$\mathbf{\check g}=-e\mathbf{A}/c+U^{-1}[\mathbf{p},U]_-$ where
$\mathbf{A}$ stands for an external vector potential. For latter
convenience, we introduce the notations: $\check b=\langle\check
b\rangle+\delta^{(1)}\check b+\delta^{(2)}\check b$, where
$\langle\check
b\rangle=\langle\check\Theta\rangle-\check\phi-\partial_t
\langle\check k\rangle$ is of the zeroth order in $\delta k$ and
$\delta \Theta$, $\delta^{(1)}\check b$ is the first order term,
and $\delta^{(2)}\check b$ is of the second order.
%\begin{gather}\label{eq:<b>}
%\langle\check b\rangle=\langle\check\Theta\rangle-\check\phi-\partial_t \langle\check k\rangle.
%\end{gather}

The particle and spin densities can be found from Eq.~\eqref{eq:S}
as
\begin{gather}\label{eq:rho_m_0}
    \begin{pmatrix}
      \rho \\
      2\mathbf{m} \\
    \end{pmatrix}=\frac i{2}\frac{\partial \ln \Z}{\partial \phi_2^\alpha}=
   - \frac{\pi\nu}2\tr(\hat\sigma_x\hat s^\alpha \langle{Q}_{t,t}\rangle)+
   2\nu \langle\check b\rangle_1^\alpha\, ,
\end{gather}
where the trace is over the Keldysh and spin spaces. Given
Eqs.~\eqref{eq:<Phi>-<Theta>} and \eqref{eq:rho_m_0}, it is
trivial to derive
\begin{gather}\label{eq:rho}
\rho=-\frac{\pi\nu\tr(\hat\sigma_x\hat s^0 \langle{\tilde
Q}\rangle)}{2(1+F_0^\rho)}-\frac{2\nu}{1+F_0^\rho}[\varphi+\partial_t\langle
k^{0}_1\rangle]\,,
\\\label{eq:m}
\mathbf m^\alpha=-\frac{\pi\nu\tr(\hat\sigma_x\hat s^\alpha
\langle{\tilde Q}\rangle)}{4(1+F_0^\sigma)}+ \frac{\nu[ g\mu_B
H_\alpha/2-\partial_t\langle
k_1^{\alpha}\rangle]}{1+F_0^\sigma}\,.
\end{gather}
%Differentiating the partition function over
%the quantum components of the source vector potential we get the expressions for particle
%($\alpha=0$) and spin ($\alpha>0$) current $
%j^\alpha=-\frac{\pi\nu}{2}
%\tr\left[\hat\sigma^x\hat s^\alpha \langle{\tilde Q}\rangle(\nabla+i\langle \mathbf{g}_1\rangle) \langle{\tilde Q}\rangle\right]$.
We emphasize that the interaction renormalizations of charge and spin densities
in Eqs.\eqref{eq:rho}-\eqref{eq:m} are in agreement with the Fermi-liquid theory~\cite{AGD}.

Although the theory~\eqref{eq:S} encodes all low-energy dynamics of the electron system, for accurate
derivation of the quantum kinetic equation it is enough to consider only a saddle-point configuration
$\underline{Q}$ for a given configuration of plasmon and gauge fields. The saddle point
(Usadel) equation is as follows
\begin{equation}\label{eq:Usadel}
D \partial_r ( \underline{Q} \partial_r\underline{Q} ) - [\partial_t -i\check b,
\underline{Q} ]_-   = 0 .
%\, \qquad \underline{Q}=\begin{pmatrix}
%  R & K \\
%  Z & A \\
%\end{pmatrix},
\end{equation}
%where $\langle Z\rangle=0$; $Z\neq0$ if there are nonzero quantum components of
%$\mathbf{\check g}_{\Sigma}$ and $\check b_{\Sigma}$ otherwise with $Z=0$, $R=-A=\hat 1$
%and in equilibrium $K=2\tanh[(\epsilon-\varphi-\partial_t \langle k^\alpha_1\rangle \hat s^\alpha)/2T]$.
A general solution of the Usadel equation can be written as
$\underline{Q}= \langle\underline{Q}\rangle+\delta\underline{Q}$
where $\langle\dots\rangle$ denotes the average over
$\delta\Theta$ fluctuations from here onwards. The term
$\delta\underline{Q}$ involves fluctuations $\delta\Theta$ and
$\delta k$ of the plasmon and gauge fields that are controlled by
the small parameter $(E_F\tau)^{-1}\ll 1$. Given
Eq.~\eqref{eq:Usadel}, it is sufficient to find
$\delta\underline{Q}$ to the second order in $\delta\Theta$ and
$\delta k$.
%We'll keep in $\delta\underline{\tilde Q}$ only terms linear and quadratic
%over the fields; these terms will  build the scattering integral.
This procedure can be significantly simplified if we demand that
$Z$-component of $\delta\underline{Q}$ does not contain linear in
$\delta \Theta,\,\delta k$ terms. It is so if
\begin{gather}\label{eq:cond1}
\delta^{(1)} b_2^\alpha-D\nabla\, \delta^{(1)}\mathbf{g}^\alpha_2=0\,.
\end{gather}
Due to the nonlinear condition $\underline{Q}^2=1$, this relation
automatically ensures that retarded ($R$) and advanced ($A$)
components of $\delta\underline{Q}$ have the same smallness as the
$Z$-component.
%Eq.\eqref{eq:cond1} relates quantum components of $\delta \check k$ and $\delta\Theta$.
We find the relation between classical components of $\delta
\Theta$ and $\delta k$ by requiring the Keldysh ($K$) component of
the Usadel equation with $\langle \underline{Q}\rangle$
substituted for $\underline{Q}$ to vanish in the linear order in
$\delta k$ and $\delta \Theta$ at $t\to t^\prime$:
%\begin{gather}\label{eq:cond_2_Fourier1}
%D\nabla\delta\mathbf{g}_1^\alpha(\omega)+\delta b_1^\alpha(\omega)=-2\omega
%[S^{-1}_\omega B_\omega]^{\alpha\beta}D\nabla\delta \mathbf{g}_2^\beta(\omega) \, .
%\end{gather}
\begin{gather}\label{eq:cond_2_Fourier1}
(D\nabla\delta\mathbf{g}+\delta b)_1^\alpha(\omega)=-2\omega [\hat
S^{-1}_\omega \hat B_\omega]^{\alpha\beta}D\nabla\delta
\mathbf{g}_2^\beta(\omega) \, .
\end{gather}
%\begin{gather}\label{eq:cond_2_Fourier1}
%(D\nabla\delta\mathbf{g}_1+\delta b_1)_\omega^\alpha=-2\omega
%[S^{-1}B]^{\alpha\beta}(D\nabla\delta \mathbf{g}_2)_\omega^\beta\, .
%\end{gather}
Here,
\begin{gather}
B^{\alpha\beta}_\omega=\frac\pi {8\omega}\int\frac{d\epsilon}{2\pi}\tr\left(
4\hat s^\alpha\hat s^\beta-\langle
K_{\epsilon_+}\rangle \hat s^\beta\langle K_{\epsilon_-}\rangle\hat s^\alpha\right)\, ,
\\
S^{\alpha\beta}_\omega=\frac \pi 4\int\frac{d\epsilon}{2\pi}\tr\left(\hat s^\alpha\langle
K_{\epsilon_+}\rangle\hat s^\beta-\hat s^\beta\langle K_{\epsilon_-}\rangle\hat s^\alpha\right)
\, .
\end{gather}
with $\epsilon_\pm=\epsilon\pm\omega/2$.
%where $\epsilon_\pm=\epsilon\pm\omega/2$,  $\langle K\rangle_{t,t'}$ is the Keldysh-component of ${\tilde \Lambda}$ and $\langle
%K\rangle_{\epsilon}=\int_t\langle K\rangle_{\tau+t/2,\tau-t/2}\exp(i\epsilon t)$.
In the absence of magnetic field, and in the equilibrium, we have
$B^{\alpha\beta}_\omega=\delta^{\alpha\beta}\coth(\omega/2T)$ and
$S^{\alpha\beta}_\omega = \omega \delta^{\alpha\beta}$. Then,
Eqs.~\eqref{eq:cond1} and \eqref{eq:cond_2_Fourier1} reduce to the
corresponding conditions of
Refs.~\cite{Kamenev-Andreev,Aleiner1,Aleiner2}. In general, we
find $\hat S_\omega=\omega \hat 1 + \hat \lambda$, where
\begin{equation}\label{eq:S_explicit}
%S^{\alpha\beta}_\omega=\omega \delta^{\alpha\beta}+
%\lambda^{\alpha\beta}, \,
\lambda^{\alpha\beta}=\frac{\pi}{4} \int\frac{d\epsilon}{2\pi} \tr\left\{
[s^\beta,s^\alpha]_{-} \langle K_\epsilon \rangle\right\}\,.
\end{equation}
%When the magnetic field is absent, in equilibrium, $B^{\alpha\beta}_\omega=\delta^{\alpha\beta}\coth(\omega/2T)$. When the conductor is spin-unpolarized, $S$ and $\lambda$ are absent and Eqs.\eqref{eq:cond1}-\eqref{eq:cond_2_Fourier1} reduce to the corresponding conditions used in Ref.\cite{Kamenev-Andreev,Aleiner1}.

Derivation of the quantum kinetic equation becomes less cumbersome
in the $\langle k\rangle$-gauge: $\langle \check b\rangle=\check
0$. Then, it implies $\delta^{(1)}\check{\mathbf{g}}=\nabla \delta
\check k$, $\delta^{(1)} \check
b=\delta\check\Theta-\partial_t\delta\check k$ and
$\delta^{(2)}\check b=i[\delta\check\Theta-\frac
12\partial_t\delta\check k, \delta \check k]_-$.
%Observables, $\rho, \mathbf{m}$, should be calculated then using
%Eq.\eqref{eq:rho_m_0}.
We mention that in the $\langle k\rangle$-gauge
$\lambda^{\alpha\beta}$ intimately related with the average spin
density: $\lambda^{\alpha\beta} = 2 i \epsilon_{\alpha\beta\gamma}
\mathbf{m}^\gamma/\nu$. {It is the presence of non-zero
$\lambda^{\alpha\beta}$, $\delta^{(2)}\check b$ and the matrix
structure of $B^{\alpha\beta}_\omega$ that strongly complicates
the derivation of the quantum kinetic equation for $\mathbf{m}\neq
0$.}

Substituting $\underline Q$ into the action~\eqref{eq:S}, we find
after expansion to the second order in $\delta \Theta$:
% We'll need the
%correlator, $\langle\delta \mathbf{g}^{\alpha}_i(rt)\,\delta
%\mathbf{g}^{\beta}_j(r't') \rangle_{\Phi}={\cal
%D}^{\alpha\beta}_{ij}(tr,t'r')/i4\nu D$. Its expression we find
%after the substitution of $\underline Q$ into the action and
% its expansion over $\delta k$ up to the second order:
\begin{gather}
i\,S[\delta\Theta]=-i\pi\nu \Tr(\delta^{(2)} b_2\, \langle
K\rangle)+i\int dt d\mathbf{r}  [\delta\Theta^\tau
\hat\Gamma^{-1}\hat\sigma_x\delta\Theta \notag \\+\nu\,
\delta^{(1)} b^\tau \hat \sigma_x \delta^{(1)} b ] +\frac{\pi\nu
D}{4} \Tr [ \delta^{(1)}\check{\mathbf{g}}, \Lambda]_-^2
.\label{ggAction}
%\mathbf{\delta\check
%g}\tilde\Lambda)-\Tr( \mathbf{\delta\check g}^2)\biggr].
\end{gather}
Given Eq.~\eqref{ggAction}, it is easy to find the 2-point
correlation function ${\cal D}^{\alpha\beta}_{ij}(\mathbf{r}
t,\mathbf{r^\prime} t^\prime) = i4\nu D \langle\delta^{(1)}
\mathbf{g}^{\alpha}_i(\mathbf{r} t)\,\delta^{(1)}
\mathbf{g}^{\beta}_j(\mathbf{r^\prime} t^\prime) \rangle$ as
%The term containing $\delta^{(2)} b_2$ is very important here [it is absent in unpolarized metals \cite{Kamenev-Andreev,Aleiner1}]; neglecting of this term brings to the essentially wrong scattering probability with the pole at Zeeman frequency. It follows that
\begin{gather}%\label{}
\hat{\mathcal{D}}_{11}(q,\omega)=\frac{
Dq^2}{Dq^2-i\omega}\left\{[(1+\hat F^{-1})D q^2-i\omega]+i
\hat\lambda\right\}^{-1}, \notag \\
\hat{\mathcal{D}}_{12}(q,\omega)= -2 i \omega
\hat{\mathcal{D}}_{11}(q,\omega) [\hat B_\omega - i (Dq^2-i\omega)
(1+\hat F^{-1})\notag \\\times  \hat S_\omega^{-1} \hat B_\omega
%\notag \\
+i (Dq^2+i\omega) \hat B_\omega \hat S_\omega^{-1} ( 1+\hat
F^{-1}) ] \hat{\mathcal{D}}_{22}(q,\omega) ,\notag \\
\hat{\mathcal{D}}_{22}(q,\omega)=[\hat{\mathcal{D}}_{11}(q,\omega)]^\dag,\quad
\hat{\mathcal{D}}_{21}(q,\omega)=0 .\label{Dprop}
\end{gather}
Substituting $\underline Q = \langle \underline{Q}\rangle
+\delta\underline{Q}$ into Eq.~\eqref{eq:Usadel}, then, expanding
its left hand side to the second order in $\delta \Theta$, and,
finally, averaging the result over the $\delta
\Theta$-fluctuations with the help of Eqs.~\eqref{eq:cond1},
\eqref{eq:cond_2_Fourier1}, and \eqref{Dprop}, we obtain the
quantum kinetic equation for the one-particle spin density matrix
$\hat f=[2\hat s^0-\langle K_\epsilon(\mathbf{r},t)\rangle]/4$:
\begin{gather}
D\triangle \hat f-\partial_\tau \hat f+\left(
e\,\mathbf{E}+\frac{\nabla (\mathbf{m} \cdot\mathbf{\hat
s})}{\nu}\right) D\partial_\epsilon \nabla \hat f=I[\hat
f],\label{QKE}
\end{gather}
where $\mathbf{E}$ stands for the electric field. It is convenient
to choose $z$ axis along $\mathbf{m}$, then $\hat f$ becomes
diagonal and the collision integral acquires the following form:
\begin{gather}%\label{}
I[f_\sigma] =\frac{16\pi}{\nu}\int\frac{d\omega}{2\pi} \Bigl \{
[P^{(\rho)}(\omega,0)+P^{(\sigma)}(\omega,0)]\,J_{\sigma,\sigma}(\epsilon,\omega)\notag
\\
+ 2
P^{(\sigma)}(\omega,|\mathbf{m}|\sigma)\,J_{\sigma,-\sigma}(\epsilon,\omega)\Bigr\}.
\label{ColInt}
\end{gather}
Here,
\begin{gather}
J_{\sigma,\sigma'}(\epsilon,\omega)=\int\frac{d\epsilon'}{2\pi}\biggl\{
(1-f_{\epsilon^\prime_+,\sigma})(1- f_{\epsilon-\omega,\sigma'})
f_{\epsilon^\prime_-,\sigma^\prime} f_{\epsilon,\sigma}  \notag
\\ - f_{\epsilon^\prime_+,\sigma} f_{\epsilon-\omega,\sigma^\prime}
(1-
f_{\epsilon^\prime_-,\sigma^\prime})(1-f_{\epsilon,\sigma})\biggr\},
\\
P^{(a)}(\omega,|\mathbf{m}|)= \sum_{\mathbf{q}} \frac{(F_0^a D q^2
|Dq^2-i\omega|^{-1})^2}{|D(1+F_0^a) q^2 -i(\omega
+\frac{2F_0^a|\mathbf{m}|}{\nu})|^2}.\notag
\end{gather}
%where $\rho(\omega,q)=D q^2/|Dq^2-i\omega|$.

Due to commutativity of the total spin with the Hamiltonian,
$I[f]$ does not lead to ``spin-flip'' processes; the spin density
evolves according to the equation: $\partial_\tau \mathbf{m} = D
\nabla^2 \mathbf{m}$. It is worthwhile mentioning, that in the
equilibrium $f_{\epsilon,\sigma}$ becomes the Fermi-Dirac
distribution function $f_F(\epsilon_\sigma)$ and $I[f_\sigma]$
vanishes identically.

The out-scattering rate can be found from $I[f_\sigma]$ by its
variation over $f_{\epsilon,\sigma}$ at the
equilibrium~\cite{Schmid,AronovAltshulerTau}. Then, from
Eqs.~\eqref{ColInt} we find $1/\tau_\textrm{out} =
1/\tau_\textrm{out}^{(s)} +
1/\tau_\textrm{out}^{(t)}(0)+2/\tau_\textrm{out}^{(t)}(|\mathbf{m}|)$
where {$1/\tau_\textrm{out}^{(s)}$ is given by the standard
expression~\cite{Schmid,Aronov-Altshuler} and}
\begin{equation}
\frac{1}{\tau_\textrm{out}^{(t)}(|\mathbf{m}|)
}=\int\frac{d\omega}{2\pi\nu}Y(\epsilon_\sigma,\omega,T)P^{(\sigma)}(\omega+2|\mathbf{m}|\sigma/\nu,
|\mathbf{m}|) \label{tauOutInt}
\end{equation}
with
$Y(\epsilon_\sigma,\omega,T)=\omega\left[\coth\frac{\omega}{2T}+
\tanh\frac{\epsilon_\sigma-\omega}{2T}\right]$. The
$P^{(a)}(\omega+2|\mathbf{m}|\sigma/\nu, |\mathbf{m}|)$ is
determined by the screened electron-electron interaction and by
the probability for electron to diffuse~\cite{Aronov-Altshuler};
in agreement with Ref.~\cite{FinkelsteinReview}, they involve
diffusion propagators for spin excitations in the presence of
$\mathbf{m}$. Evaluating integrals over momentum and frequencies
in Eqs.~\eqref{tauOutInt} for $T=0$ and $d=3$ we obtain the
result~\eqref{tauOutd3}. As known very well in lower dimensions
and non-zero temperature one should evaluate the integrals in
Eq.~\eqref{tauOutInt} self-consistently~\cite{Blanter}; we present
detailed results {for $T>0$ and $d=1,2,3$}
elsewhere~\cite{Future}.

In summary, we derived the quantum kinetic equation that describes
the energy relaxation due to electron-electron interaction in
disordered electron systems in the presence of non-zero spin
polarization. We found that the $T=0$ rate of electron-electron
collisions is non-monotonic function of the electron
magnetization. It can be used for decoupling electron degrees of
freedom from the environment.

We would like to thank V. Ryazanov for drawing our attention to
the problem and B.L. Altshuler for pointing out the
Ref.~\cite{Aleiner2}. The research was funded in part by RFBR, the
Russian Ministry of Education and Science, the Council for grants
of the President of Russian Federation, Russian Science Support
Foundation, Dynasty Foundation, the Program of RAS ``Quantum
Macrophysics'', CRDF and NWO.

\end{document}